\documentclass[10pt]{IEEETran}

\def \be {\begin{equation}}
\def \ee {\end{equation}}

\def \nn {\nonumber}

\usepackage{ifpdf}

\usepackage[dvips]{graphicx}
\usepackage{color}

\usepackage[cmex10]{amsmath}
\usepackage{amssymb}

\begin{document}
%
\title{\LARGE \bf Dynamic Feedback for Consensus of Networked Lagrangian Systems}
%
%
%

\author{Hanlei~Wang 
\thanks{The author is with the Science and Technology on Space Intelligent Control Laboratory,
Beijing Institute of Control Engineering,
Beijing 100094, China (e-mail: hlwang.bice@gmail.com).
%
%
%
%
}
}
\maketitle

\begin{abstract}
This paper investigates the consensus problem of multiple uncertain Lagrangian systems. Due to the discontinuity resulted from the switching topology, achieving consensus in the context of uncertain Lagrangian systems is challenging. We propose a new adaptive controller based on dynamic feedback to resolve this problem and additionally propose a new analysis tool for rigorously demonstrating the stability and convergence of the networked systems. The new introduced analysis tool is referred to as uniform integral-$\mathcal L_p$ stability, which is motivated for addressing integral-input-output properties of linear time-varying systems. It is then shown that the consensus errors between the systems converge to zero so long as the union of the graphs contains a directed spanning tree. It is also shown that the proposed controller enjoys the robustness with respect to constant communication delays. The performance of the proposed adaptive controllers is shown by numerical simulations.
\end{abstract}


\section{Introduction}

Controlled collective behaviors of networked systems are of particular interest in recent years due in part to their potential applications in many engineering problems (e.g., cooperative monitoring by multiple unmanned aerial vehicles (UAVs) and synchronized manipulation by multiple robots). To serve this purpose, many distributed controllers have been proposed to resolve the fundamental issues in maintaining the collective motion of networked systems, e.g., interaction topology \cite{Olfati-Saber2004_TAC,Ren2005_TAC,Liu2014_SCL,Liu2014_JFI,Cai2014_IJGS}, communication delays \cite{Chopra2006,Tian2008_TAC,Lestas2010_AUT,Nuno2011_TAC,Abdessameud2014_TAC,Wang2014_TAC}, and model uncertainties \cite{Nuno2011_TAC,Abdessameud2014_TAC,Wang2014_TAC,Wang2013b_AUT}; see also \cite{Knorn2016_TCNS} and the references therein.

It might often be the case that the network issues are intertwined with the dynamics of agents (e.g., nonlinearities and uncertainties); for instance, the collective control of multiple Lagrangian systems \cite{Chopra2006,Ren2009_IJC,Liu2013_AUT,Nuno2011_TAC,Min2012_SCL,Mei2012_AUT,Mei2013_AUT,Wang2013_TAC,Abdessameud2014_TAC,Wang2014_TAC,Ghapani2016_AUT,Nuno2017_TCST}.
In particular, dynamic feedback is proposed for achieving the second-order consensus \cite{Wang2013b_AUT,Wang2015_IJC}, flocking \cite{Ghapani2016_AUT}, or robustness with respect to communication delays \cite{Abdessameud2014_TAC,Nuno2017_TCST}; new tools are also introduced to resolve the related new issues, especially in the case of directed topology (e.g., iBIBO-stability-based analysis in \cite{Wang2014_TAC,Wang2015_IJC} and small-gain-based analysis in \cite{Abdessameud2014_TAC}). The issue of switching topology in the context of multiple Lagrangian systems is addressed in \cite{Chopra2006,Liu2014_SCL,Liu2014_JFI,Cai2014_IJGS,LiuYC2015_JFI}, either taking into consideration the model uncertainties (e.g., \cite{Liu2014_SCL,Cai2014_IJGS,LiuYC2015_JFI}) or assuming the exact knowledge of the system model (e.g., \cite{Chopra2006,Liu2014_JFI}). These control schemes for switching topology can be grouped into two categories: passivity-based scheme (e.g., \cite{Liu2014_SCL,Liu2014_JFI,LiuYC2015_JFI}) and dynamic-compensator-based scheme (e.g., \cite{Cai2014_IJGS}). The passivity-based adaptive scheme, as stated in \cite{Nuno2010_AUT,Nuno2011_TAC}, gives rise to the consequence that the positions of the systems converge to the origin in the presence of gravitational torques. The dynamic-compensator-based scheme in \cite{Cai2014_IJGS}, by separating the design of the network coupling and that of the controller design for each system, avoids this issue but this kind of distributed-observer-based control relies on the communication of artificially produced quantities (not physical quantities such as positions or velocities); in addition, this scheme is not manipulable in the sense of \cite{Wang2017_arXiv}, i.e., the consensus behavior cannot be maintained in the case of an external human physical input (mainly due to the fact that the network coupling dynamics acts as a reference command and it does not respond to any physical evolution of the system except for the leader). In this sense, the consensus problem for multiple uncertain Lagrangian systems with directed switching topology in the case of only using physically coupled action is still unresolved. Using only physically coupled action mimics the collective behaviors in nature, and meanwhile implies the cost efficiency since the mutual communication between the neighboring systems is not required. Even in the case of acquiring relative position and velocity information by communication, the use of physically coupled action is preferable for its strong manipulability in the sense of \cite{Wang2017_arXiv}.

In this paper, we propose an adaptive controller based on dynamic feedback for realizing consensus of multiple Lagrangian systems. To show the convergence of the system under switching topology, we establish several new input-output properties concerning linear time-varying systems (which is resulted from the switching topology). These new input-output properties are referred to as uniform integral-${\mathcal L}_p$ stability since it involves linear time-varying systems and describes the relation between the integral of the input and the output, in contrast with the standard $\mathcal L_p$ stability concerning linear time-invariant (LTI) systems (see, e.g., \cite{Desoer1975_Book,Ioannou1996_Book}) and also with the integral-$\mathcal L_p$ stability concerning marginally stable LTI systems (see, e.g., \cite{Wang2017_arXiv,Wang2015_IJC}). By the introduced new tools, the convergence of the consensus errors is rigorously shown under the very mild condition that the union of the graphs contains a directed spanning tree. The proposed controller only uses the physically coupled action between the neighboring systems, in contrast with \cite{Cai2014_IJGS}, and in addition the proposed controller ensures that the positions of the systems converge to a common value (typically nonzero), in contrast with the passivity-based adaptive schemes in, e.g., \cite{Liu2014_SCL,Liu2014_JFI,LiuYC2015_JFI} (the consensus equilibrium of the system under these passivity-based adaptive schemes is the origin in the presence of gravitational torques). The condition that the possible interaction topologies are balanced-like or regular (see, e.g., \cite{Liu2014_SCL,Liu2014_JFI,LiuYC2015_JFI}) is no longer required due to the proposed adaptive controller based on dynamic feedback and the proposed new analysis tool.

The adaptive controllers in \cite{Wang2013b_AUT,Wang2015_IJC,Ghapani2016_AUT,Abdessameud2014_TAC} also rely on dynamic feedback yet the interaction topology is assumed to be invariant. Our result considers the case of switching topology and in particular resolves the issues concerning discontinuity and time-varying nature of the system by resorting to dynamic feedback and a new analysis tool (i.e., the uniform integral-$\mathcal L_p$ stability). We also show by the uniform integral-$\mathcal L_p$ stability tool that the proposed controller is valid under both the switching topology and constant communication delays provided that the communication delays are bounded, and the communication delays are not required to be uniform or exactly known. 

\section{Preliminaries}

\subsection{Graph Theory}

Let us give a brief introduction of the graph theory \cite{Godsil2001_Book,Olfati-Saber2004_TAC,Ren2005_TAC,Ren2008_Book} in the context that $n$ Lagrangian systems are involved. As is commonly done, we employ a directed graph ${\mathcal G}=({\mathcal V},{\mathcal E})$ to describe the interaction topology among the $n$ systems where ${\mathcal V}=\left\{1,\dots,n\right\}$ is the vertex set that denotes the collection of the $n$ systems and ${\mathcal E}\subseteq {\mathcal V}\times {\mathcal V}$ is the edge set that denotes the information interaction among the $n$ systems. The set of neighbors of system $i$ is denoted by $\mathcal{N}_i=\left\{j|(i,j)\in \mathcal{E}\right\}$. A graph is said to have a directed spanning tree if there is a vertex $k_0\in \mathcal{V}$ such that any other vertex of the graph has a directed path to $k_0$. The weighted adjacency matrix $\mathcal{W}=\left[w_{ij}\right]$ associated with the graph $\mathcal G$ is defined as $w_{ij}>0$ if $j\in \mathcal{N}_i$, and $w_{ij}=0$ otherwise. Furthermore, it is assumed that $w_{ii}=0$, $\forall i=1,2,\dots,n$. The Laplacian matrix $\mathcal{L}_w=\left[\ell_{w,ij}\right]$ associated with the graph $\mathcal G$ is defined as $\ell_{w,ij}=\Sigma_{k=1}^n w_{ik}$ if $i=j$, and $\ell_{w,ij}=-w_{ij}$ otherwise. Several basic properties concerning the Laplacian matrix $\mathcal{L}_w$ can be described by the following lemma.

\emph{Lemma 1 (\cite{Lin2005_TAC,Ren2005_TAC,Ren2008_Book}):} If $\mathcal{L}_w$ is associated with a directed graph containing a directed spanning tree, then
 \begin{enumerate}
 \item $\mathcal{L}_w$ has a simple zero eigenvalue, and all other eigenvalues of $\mathcal{L}_w$ have positive real parts;
 \item $\mathcal{L}_w$ has a right eigenvector $1_n=\left[1, \dots , 1\right]^T$ and a nonnegative left eigenvector $\gamma=\left[\gamma_1, \dots, \gamma_n\right]^T$ satisfying $\Sigma_{k=1}^n \gamma_k=1$ associated with its zero eigenvalue, i.e., $\mathcal{L}_w 1_n=0$ and $\gamma^T \mathcal{L}_w=0$.
\end{enumerate}
In the case of switching topology, the interaction graphs among the systems are dynamically changing. Denote by ${\mathcal G}_S=\{\mathcal G_1,\dots,\mathcal G_{n_s}\}$ the set of all possible interaction graphs among the $n$ systems, and these graphs share the same vertex set $\mathcal V$, but their edge sets may be different. The union of a collection of graphs $G_{i_1}, \dots,G_{i_s}$ with $i_s\le n_s$ is a graph with vertex set given by $\mathcal V$ and edge set given by the union of the edge sets of $G_{i_1}, \dots,G_{i_s}$. 

\subsection{Equations of Motion of Lagrangian Systems}

The equations of motion of the $i$-th $m$-DOF (degree-of-freedom) Lagrangian system can be written as \cite{Slotine1991_Book,Spong2006_Book}
\be
\label{eq:1}
M_i(q_i)\ddot q_i+C_i(q_i,\dot q_i)\dot q_i+g_i(q_i)=\tau_i
\ee
where $q_i\in R^m$ is the generalized position (or configuration), $M_i(q_i)\in R^{m\times m}$ is the inertia matrix, $C_i(q_i,\dot q_i)\in R^{m\times m}$ is the Coriolis and centrifugal matrix, $g_i(q_i)\in R^m$ is the gravitational torque, and $\tau_i\in R^m$ is the exerted control torque. Three well-known properties associated with the dynamics (\ref{eq:1}) are listed as follows.

\emph{Property 1 (\cite{Slotine1991_Book,Spong2006_Book}):} The inertia matrix $M_i(q_i)$ is symmetric and uniformly positive definite.

\emph{Property 2 (\cite{Slotine1991_Book,Spong2006_Book}):} With a suitable choice of $C_i(q_i,\dot q_i)$, the matrix $\dot M_i(q_i)-2C_i(q_i,\dot q_i)$ is skew-symmetric.

\emph{Property 3 (\cite{Slotine1991_Book,Spong2006_Book}):} The dynamics (\ref{eq:1}) depends linearly on a constant dynamic parameter vector $\vartheta_i$, which yields
\be
M_i(q_i)\dot\zeta+C_i(q_i,\dot q_i)\zeta+g_i(q_i)=Y_i(q_i,\dot q_i,\zeta,\dot \zeta)\vartheta_i
\ee
where $\zeta\in R^m$ is a differentiable vector, $\dot \zeta$ is the derivative of $\zeta$, and $Y_i(q_i,\dot q_i,\zeta,\dot \zeta)$ is the regressor matrix.

\section{Consensus With Switching Topology}

In this section, we develop an adaptive controller to realize consensus of the $n$ Lagrangian systems with switching topology. The control objective is to ensure that $q_i-q_j\to 0$ and $\dot q_i\to0$ as $t\to\infty$, $\forall i,j=1,\dots,n$. To this end, introduce the following dynamic system
\be
\label{eq:3}
\dot z_i=-{\alpha \dot q_i}-\Sigma_{j\in{\mathcal N}_i(t)}w_{ij}(t)[(\dot q_i+\alpha q_i)-(\dot q_j+\alpha q_j)]
\ee
with $\alpha$ being a positive design constant and $z_i\in R^m$, and define
\be
\label{eq:4}
s_i=\dot q_i-z_i.
\ee
The adaptive controller is given as
\be
\label{eq:5}
\begin{cases}
\tau_i=-K_i s_i+Y_i(q_i,\dot q_i,z_i,\dot z_i)\hat\vartheta_i\\
\dot{\hat \vartheta}_i=-\Gamma_i Y_i^T(q_i,\dot q_i,z_i,\dot z_i)s_i
\end{cases}
\ee
where $K_i$ is a symmetric positive definite matrix and $\hat\vartheta_i$ is the estimate of $\vartheta_i$. The adaptive controller given by (\ref{eq:5}) leads to the following dynamics for describing the behavior of the $i$-th system
\be
\label{eq:6}
\begin{cases}
\dot q_i=z_i+s_i\\
\dot z_i=-{\alpha \dot q_i}-\Sigma_{j\in{\mathcal N}_i(t)}w_{ij}(t)[(\dot q_i+\alpha q_i)-(\dot q_j+\alpha q_j)]\\
M_i(q_i)\dot s_i+C_i(q_i,\dot q_i)s_i=-K_i s_i+Y_i(q_i,\dot q_i,z_i,\dot z_i)\Delta \vartheta_i\\
\dot{\hat \vartheta}_i=-\Gamma_i Y_i^T(q_i,\dot q_i,z_i,\dot z_i)s_i
\end{cases}
\ee
where $\Delta \vartheta_i=\hat\vartheta_i-\vartheta_i$.

\emph{Remark 1:} The switching interaction graph introduces discontinuous quantities [e.g., the adjacency weight $w_{ij}(t)$]. The existing adaptive controllers concerning the static consensus problem for Lagrangian systems (e.g., \cite{Nuno2011_TAC,Mei2012_AUT,Wang2014_TAC}) is based on static feedback in terms of the neighboring position and velocity information, and this, unfortunately, would involve the differentiation of the discontinuous adjacency weight among the systems in the case that the interaction graph is switching. Here by resorting to dynamic feedback (i.e., by dynamically generating a new vector $z_i$), this undesirable problem is resolved and the control torque no longer involves the differentiation of the discontinuous adjacency weight. On the other hand, the newly encountered stability issues of the system under this dynamic-feedback-based design also motivates the introduction of a new analysis tool (as is discussed later).

\emph{Theorem 1:} Let $t_0,t_1,t_2,\dots$ denote a series of time instants at which the interaction graph switches and these instants satisfy that $0=t_0<t_1<t_2<\dots$ and that $T_D\le t_{k+1}-t_k<T_0$, $\forall k=0,1,\dots$, for some positive constants $T_D$ and $T_0$. If there exists an infinite number of uniformly bounded intervals $[t_{i_p},t_{i_{p+1}})$, $p=1,2,\dots$ with $t_{i_1}=t_0$ satisfying the property that the union of the interaction graphs in each interval contains a directed spanning tree, then the adaptive controller given by (\ref{eq:5}) ensures the consensus of the $n$ systems, i.e., $q_i-q_j\to 0$ and $\dot q_i\to 0$ as $t\to \infty$, $\forall i,j=1,\dots,n$.

Before proving Theorem 1, we first present the following proposition for describing the integral-input-output properties for linear time-varying systems.

\emph{Proposition 1:} Consider a linear time-varying system with an external input
\be
\label{eq:7}
\dot y=A(t) y+ u
\ee
where $y$ is the output, $A(t)$ is the system coefficient matrix and is uniformly bounded, and $u$ acts as the external input. If the linear time-varying system is uniformly asymptotically stable, the system (\ref{eq:7}) is uniformly integral-bounded-input bounded-output stable, i.e., if  $\int_0^t u(\sigma)d\sigma\in{\mathcal L}_\infty$, then $y\in{\mathcal L}_\infty$. In addition, if $\int_0^t u(\sigma)d\sigma+c\in{\mathcal L}_p$ with $c$ being an arbitrary constant, then $y\in{\mathcal L}_p$, $p\in[1,\infty)$.

\emph{Proof:} Let $y^\ast=y-\int_0^t u(\sigma)d\sigma$ and $u^\ast=\int_0^t u(\sigma)d\sigma$, and we then have that
\be
\label{eq:8}
\dot y^\ast=A(t)y^\ast+A(t)u^\ast
\ee
In the case that the linear time-varying system is uniformly asymptotically stable, then the perturbed linear time-varying system is uniformly bounded-input bounded-output stable, according to the standard linear system theory (see, e.g., \cite{Rugh1996_Book}). Hence we obtain that $y^\ast\in {\mathcal L}_\infty$, which immediately leads us to obtain that $y\in{\mathcal L}_\infty$.

For $p\in[1,\infty)$, we first consider the case that $c=0$. The uniform asymptotic stability of the time-varying system $\dot y^\ast=A(t)y^\ast$ implies that there exist positive constants $\ell_1$ and $\ell_2$ such that \cite{Rugh1996_Book}
\be
\label{eq:9}
|\Phi(t,t_0)|\le\ell_1 e^{-\ell_2 (t-t_0)}, \forall t\ge t_0
\ee
where $\Phi(t,t_0)$ denotes the transition matrix of the time-varying system. As is known, the solution of (\ref{eq:8}) can be written as
\be
y^\ast(t)=\underbrace{\Phi(t,0)y^\ast(0)}_{y^\ast_1}+\underbrace{\int_0^t \Phi(t,\sigma)A(\sigma)u^\ast(\sigma)d\sigma}_{y^\ast_2}.
\ee
It is apparent that the signal $y^\ast_1\in{\mathcal L}_p$ since it uniformly exponentially converges to zero. Consider now the variable $y^\ast_2$. In the case that $p=1$, we have that
\begin{align}
\|y_2^\ast\|_1=&\int_0^\infty|y_2^\ast(\sigma)|d\sigma\nn\\
\le &\int_0^\infty\int_0^\sigma|\Phi(\sigma,r)||A(r)||u^\ast(r)|drd\sigma\nn\\
\le & \sup_{t\in[0,\infty)} |A(t)| \int_0^\infty\int_0^\sigma|\Phi(\sigma,r)||u^\ast(r)|drd\sigma\nn\\
=& \sup_{t\in[0,\infty)} |A(t)| \int_0^\infty\int_r^\infty|\Phi(\sigma,r)||u^\ast(r)|d\sigma dr\nn\\
\le &\sup_{t\in[0,\infty)} |A(t)| \int_0^\infty\int_0^\infty|\Phi(\sigma,0)|d\sigma |u^\ast(r)| dr\nn\\
=&\sup_{t\in[0,\infty)} |A(t)| \int_0^\infty|\Phi(\sigma,0)|d\sigma\int_0^\infty |u^\ast(r)| dr\nn\\
=& \sup_{t\in[0,\infty)} |A(t)| \|\Phi\|_1\|u^\ast\|_1
\end{align}
with $\|\Phi\|_1$ being defined as $\|\Phi\|_1=\int_0^\infty\Phi(\sigma,0)d\sigma$, which satisfies that $\|\Phi\|_1\le \ell_1/\ell_2$ due to (\ref{eq:9}). This immediately leads to the conclusion that $y_2^\ast\in{\mathcal L}_1$, and therefore $y^\ast\in {\mathcal L}_1$.
In the case that $p>1$, introduce a constant $l$ such that $1/p+1/l=1$. Then
\begin{align}
&\|y_2^\ast\|_p=\left[\int_0^\infty|y_2^\ast(\sigma)|^p d\sigma\right]^{\frac{1}{p}}\nn\\
\le &\left\{\int_0^\infty\left[\int_0^\sigma|\Phi(\sigma,r)||A(r)||u^\ast(r)|dr\right]^pd\sigma\right\}^{\frac{1}{p}}\nn\\
\le & \sup_{t\in[0,\infty)} |A(t)|\left\{ \int_0^\infty\left[\int_0^\sigma|\Phi(\sigma,r)|^{\frac{1}{p}}|u^\ast(r)|\Phi(\sigma,r)|^{\frac{1}{l}}dr\right]^pd\sigma\right\}^{\frac{1}{p}}\nn\\
\le & \sup_{t\in[0,\infty)} |A(t)|\left\{ \int_0^\infty \int_0^\sigma|\Phi(\sigma,r)||u^\ast(r)|^p dr \right.\nn\\
&\left.\times
\left[\int_0^\sigma |\Phi(\sigma,r)|dr \right]^{\frac{p}{l}}d\sigma\right\}^{\frac{1}{p}}\nn\\
\le& \sup_{t\in[0,\infty)} |A(t)|\|\Phi\|_1^{\frac{1}{l}} \left\{ \int_0^\infty \int_0^\sigma|\Phi(\sigma,r)||u^\ast(r)|^p dr d\sigma\right\}^{\frac{1}{p}}\nn\\
=& \sup_{t\in[0,\infty)} |A(t)|\|\Phi\|_1^{\frac{1}{l}} \left\{ \int_0^\infty \int_r^\infty|\Phi(\sigma,r)||u^\ast(r)|^p d\sigma dr\right\}^{\frac{1}{p}}\nn\\
\le& \sup_{t\in[0,\infty)} |A(t)|\|\Phi\|_1^{\frac{1}{l}+\frac{1}{p}}\|u^\ast\|_p=\sup_{t\in[0,\infty)} |A(t)|\|\Phi\|_1 \|u^\ast\|_p
\end{align}
which gives rise to the consequence that $y_2^\ast\in{\mathcal L}_p$, and hence $y^\ast\in{\mathcal L}_p$.

In the case $c\ne 0$, we can redefine $y^\ast=y-\int_0^t u(\sigma)d\sigma-c$ and $u^\ast=\int_0^t u(\sigma)d\sigma+c$, and equation (8) with this redefinition still holds. Therefore, the same conclusion follows. \hfill {\small $\blacksquare$}

\emph{Remark 2:} The ${\mathcal L}_p$ stability described in Proposition 1 as well as the proof extends the results for linear time-invariant systems in \cite[p.~59, p.~240, p.~241]{Desoer1975_Book}. An important difference is that the ${\mathcal L}_p$ stability here is concerning the relation between the output and integral of the input for linear time-varying systems (in contrast with \cite{Desoer1975_Book}), and we thus refer to these integral-input-output properties as uniform integral-${\mathcal L}_p$ stability.

The uniform $\mathcal L_p$ stability in terms of the relation between the output $y$ and input $u$ of (\ref{eq:7}) can be similarly derived as the uniform integral-$\mathcal L_p$ stability.

\emph{Proposition 2:} Suppose that the linear time-varying system (\ref{eq:7}) with $u=0$ is uniformly asymptotically stable and $A(t)$ is uniformly bounded. Then
\begin{enumerate}
\item if $u\in{\mathcal L}_\infty$, $y\in{\mathcal L}_\infty$;
\item if $u\in{\mathcal L}_p$, $p\in[1,\infty)$, $y\in{\mathcal L}_p\cap {\mathcal L}_\infty$, $\dot y\in{\mathcal L}_p$, and $y\to 0$ as $t\to\infty$.
\end{enumerate}

The uniform $\mathcal L_\infty$ stability described in Proposition 2 is equivalent to the uniform bounded-input bounded-output stability in \cite{Rugh1996_Book}, and the uniform $\mathcal L_p$ stability here extends the $\mathcal L_p$ stability given in \cite{Desoer1975_Book,Ioannou1996_Book} to the case of linear time-varying systems.

\emph{Proof of Theorem 1:} Following the typical practice (see, e.g., \cite{Ortega1989_AUT,Slotine1987_IJRR}), we consider the  Lyapunov-like function candidate $
V_i=(1/2)s_i^T M_i(q_i)s_i+(1/2)\Delta \vartheta_i^T\Gamma_i^{-1}\Delta\vartheta_i
$ and its derivative along the trajectories of the system can be written as
$
\dot V_i=-s_i^T K_i s_i\le 0,
$
which gives that $s_i\in{\mathcal L}_2\cap {\mathcal L}_\infty$ and $\hat\vartheta_i\in {\mathcal L}_\infty$, $\forall i$. From the first two subsystems of (\ref{eq:6}), we obtain that
\be
\label{eq:13}
\ddot q_i=-\alpha\dot q_i -\Sigma_{j\in{\mathcal N}_i(t)}w_{ij}(t)[(\dot q_i+\alpha q_i)-(\dot q_j+\alpha q_j)]+\dot s_i.
\ee
To this end, define a sliding vector (the same as \cite{Chopra2006})
\be
\label{eq:14}
\xi_i=\dot q_i+\alpha q_i
\ee
and by this vector, we can rewrite (\ref{eq:13}) as
\be
\label{eq:15}
\dot \xi_i=-\Sigma_{j\in{\mathcal N}_i(t)}w_{ij}(t)(\xi_i-\xi_j)+\dot s_i
.\ee
We can write (\ref{eq:15}) in matrix form as
\be
\dot \xi=-[{\mathcal L}_w(t)\otimes I_m]\xi+\dot s
\ee
where $\xi=[\xi_1^T,\dots,\xi_n^T]^T$ and $s=[s_1^T,\dots,s_n^T]^T$, $\otimes$ denotes the Kronecker product \cite{Brewer1978_TCS}, and the Laplacian matrix ${\mathcal L}_w(t)$ is switching (not continuous) due to the switching of the interaction topology. Let $\xi_{E,i}=\xi_i-\xi_{i+1}$, $i=1,\dots,n-1$ and $\xi_E=[\xi_{E,1}^T,\dots,\xi_{E,n-1}^T]^T$, we then obtain
\be
\label{eq:17}
\dot \xi_E=-\Omega(t)\xi_E+\dot {s}_E
\ee
where $\Omega(t)$ is a time-varying matrix (due to the switching of the interaction graph) and $s_E=[s_1^T-s_2^T,\dots,s_{n-1}^T-s_n^T]^T\in{\mathcal L}_2\cap{\mathcal L}_\infty$. According to \cite[p.~48, p.~49]{Ren2008_Book}, the linear time-varying system
\be
\dot \xi_E=-\Omega(t)\xi_E
\ee
is uniformly asymptotically stable. Then from Proposition 1, we obtain from (\ref{eq:17}) that $\xi_E\in{\mathcal L}_2\cap{\mathcal L}_\infty$ since $\int_0^t \dot s_E(\sigma)d\sigma+s_E(0)=s_E\in{\mathcal L}_2\cap{\mathcal L}_\infty$. From (\ref{eq:14}), we obtain that
\be
\xi_E=\dot q_E+\alpha q_E
\ee
with $q_E=[q_1^T-q_2^T,\dots,q_{n-1}^T-q_n^T]^T$, and this immediately leads to the result that $q_E\in{\mathcal L}_2\cap{\mathcal L}_\infty$, $\dot q_E\in{\mathcal L}_2\cap{\mathcal L}_\infty$, and $q_E\to 0$ as $t\to\infty$ from the input-output properties of strictly proper and exponentially stale linear systems \cite[p.~59]{Desoer1975_Book}. Using (\ref{eq:4}), equation (\ref{eq:3}) can be rewritten as
\be
\dot z_i=-\alpha z_i\underbrace{-\Sigma_{j\in{\mathcal N}_i(t)}w_{ij}(t)[(\dot q_i+\alpha q_i)-(\dot q_j+\alpha q_j)]-\alpha s_i}_{\Delta_{s,i}}
\ee
and considering the fact that $\Delta _{s,i}\in{\mathcal L}_2\cap{\mathcal L}_\infty$, we obtain that $z_i\in{\mathcal L}_2\cap{\mathcal L}_\infty$, $\dot z_i\in{\mathcal L}_2\cap{\mathcal L}_\infty$, and $z_i\to 0$ as $t\to\infty$ from the input-output properties of exponentially stable and strictly proper linear systems \cite[p.~59]{Desoer1975_Book}, $\forall i$. Therefore, $\dot q_i=z_i+s_i\in{\mathcal L}_2\cap{\mathcal L}_\infty$, $\forall i$. From the third subsystem of (\ref{eq:6}) and using Property 1, we obtain that $\dot s_i\in{\mathcal L}_\infty$ and thus $s_i$ is uniformly continuous, $\forall i$. From the properties of square-integrable and uniformly continuous functions \cite[p.~232]{Desoer1975_Book}, we obtain that $s_i\to 0$ as $t\to\infty$, $\forall i$. Hence, $\dot q_i\to 0$ as $t\to\infty$, $\forall i$. From (\ref{eq:13}), we obtain that $\ddot q_i\in{\mathcal L}_\infty$, $\forall i$.
\hfill{\small {$\blacksquare $}}

\emph{Remark 3:} An important portion in the proof of Theorem 1 is to analyze the system (\ref{eq:17}) with $s_E\in{\mathcal L}_2\cap{\mathcal L}_\infty$ (but we do not know properties directly concerning $\dot s_E$). This is quite different from the standard setting of input-output properties of dynamical systems (see, e.g., \cite{Desoer1975_Book,Sontag1998_SCL}), which involves the relation between the input and output/state. Here we only know some properties of the integral of the input $\dot s_E$, i.e., $\int_0^t \dot s_E(\sigma)d\sigma+s_E(0)=s_E\in{\mathcal L}_2\cap{\mathcal L}_\infty$. The uniform bounded-input bounded-output property of (\ref{eq:17}) with $\dot s_E$ as the input and $\xi_E$ as the output is shown in \cite[p.~48, p.~49]{Ren2008_Book}, but this, however, is not the case here.


\section{Consensus With Communication Delays and Switching Topology}

In this section, we consider the case of existence of communication delays and the delays are assumed to be constant and bounded (not required to be known exactly).

We start by considering the simplified case that the topology is fixed and communication delays exist among the $n$ systems. In this context, we define the vector $z_i$ by
\be
\label{eq:21}
\dot z_i=-\alpha \dot q_i-\Sigma_{j\in \mathcal N_i}w_{ij}[\xi_i-\xi_j(t-T_{ij})]
\ee
where $\xi_i$ is defined as (\ref{eq:14}), and $T_{ij}$ is the communication delay from system $j$ to system $i$. The adaptive controller remains the same as (\ref{eq:5}).

\emph{Theorem 2:} The adaptive controller (\ref{eq:5}) with $z_i$ being given by (\ref{eq:21}) ensures the consensus of the $n$ Lagrangian systems provided that the interaction graph contains a directed spanning tree, i.e., $q_i-q_j\to 0$ and $\dot q_i\to 0$ as $t\to\infty$, $\forall i,j=1,\dots,n$.

\emph{Proof:} Most of the proof is similar to that of Theorem 1. The main difference is that the interconnection system becomes [unlike (\ref{eq:15})]
\be
\dot \xi_i=-\Sigma_{j\in\mathcal N_i}w_{ij}[\xi_i-\xi_j(t-T_{ij})]+\dot s_i, i=1,\dots,n.
\ee
From the above system, we can obtain
\be
\dot \xi_E={\mathcal F}(\xi_E)+\dot s_E
\ee
and according to the existing literature (see, e.g., \cite{Tian2008_TAC}), the linear system
\be
\dot \xi_E={\mathcal F}(\xi_E)
\ee
is asymptotically stable and thus exponentially stable from the standard linear system theory. Then from Proposition 1, we obtain that $\xi_E\in{\mathcal L}_2\cap{\mathcal L}_\infty$. Then it can be shown by following similar procedures as in the proof of Theorem 1 that $q_i-q_j\to 0$ and $\dot q_i\to 0$ as $t\to\infty$, $\forall i,j$. \hfill{\small {$\blacksquare $}}

\emph{Remark 4:} A direct benefit of the adaptive controller here is the reduction of the communicated information in comparison with \cite{Nuno2011_TAC,Wang2014_TAC}, and only the composite of the position and velocity information (i.e., $\xi_i$, $i=1,\dots,n$) needs to be shared among the systems while both the position and velocity information are required to be shared in \cite{Nuno2011_TAC,Wang2014_TAC}.

We next consider the consensus with both the communication delays and switching topology, and we define
\be
\label{eq:25}
\dot z_i=-\alpha \dot q_i-\Sigma_{j\in \mathcal N_i(t)}w_{ij}(t)[\xi_i-\xi_j(t-T_{ij})].
\ee
In comparison with the fixed topology case, $\mathcal N_i(t)$ and $w_{ij}(t)$ are time-varying rather than time invariant as (\ref{eq:21}).

\emph{Theorem 3:} If there exists an infinite number of uniformly bounded intervals $[t_{i_p},t_{i_{p+1}})$, $p=1,2,\dots$ with $t_{i_1}=t_0$ satisfying the property that the union of the interaction graphs in each interval contains a directed spanning tree, then the adaptive controller given by (\ref{eq:5}) with $z_i$ being defined by (\ref{eq:25}) ensures the consensus of the $n$ systems, i.e., $q_i-q_j\to 0$ and $\dot q_i\to 0$ as $t\to \infty$, $\forall i,j=1,\dots,n$.

The proof of Theorem 3 relies on the study of the following interconnection system
\be
\dot \xi_i=-\Sigma_{j\in\mathcal N_i(t)}w_{ij}(t)[\xi_i-\xi_j(t-T_{ij})]+\dot s_i, i=1,\dots,n,
\ee
or the stability properties of its reduced version
\be
\label{eq:27}
\dot \xi_i=-\Sigma_{j\in\mathcal N_i(t)}w_{ij}(t)[\xi_i-\xi_j(t-T_{ij})], i=1,\dots,n.
\ee
To this end, we recall the analysis approach in \cite{Moreau2004_CDC}. Specifically consider the following nonnegative Lyapunov functional
\begin{align}
V_k^\ast(t)=&\max_{\sigma\in[t-T_{\max},t]}\{\xi_1^{(k)}(\sigma),\dots,\xi_n^{(k)}(\sigma)\}\nn\\
&-\min_{\sigma\in[t-T_{\max},t]}\{\xi_1^{(k)}(\sigma),\dots,\xi_n^{(k)}(\sigma)\}, k=1,\dots,m
\end{align}
where $T_{\max}$ is the upper bound of the communication delays among the $n$ systems and $\xi_i^{(k)}$ is the $k$-th entry of $\xi_i$, $\forall i$, and the exponential stability of (\ref{eq:27}) can be derived (see \cite{Moreau2004_CDC}). Then using Proposition 1, we can complete the proof of Theorem 3.

\section{Simulation Results}

Consider a network consisting of six two-DOF robots, and the interaction graph of the six robots randomly switches among the ones shown in Fig. 1. Physical parameters of the robots are not listed here for saving space. The sampling period is chosen as 5 ms. The interaction graph randomly switches among the three graphs in Fig. 1 every 50 ms according to the uniform distribution.

 The initial joint positions of the robots are set as $q_1(0)=[\pi/6,\pi/3]^T$, $q_2(0)=[-\pi/6,\pi/6]^T$, $q_3(0)=[-\pi/2,\pi/2]^T$, $q_4(0)=[\pi/3,-\pi/6]^T$, $q_5(0)=[-2\pi/3,-2\pi/3]^T$, and $q_6(0)=[\pi/2,-\pi/2]^T$. The initial joint velocities of the robots are set as $\dot q_i(0)=[0,0]^T$, $i=1,\dots,6$. The controller parameters are chosen as $K_i=30.0 I_2$, $\alpha=3.0$, and $\Gamma_i=6.0 I_3$, $i=1,\dots,6$. The initial values of $z_i$, $i=1,\dots,6$ are set as $z_i(0)=[0,0]^T$. The adjacency weights are set as $w_{ij}(t)=1.0$ if $j\in{\mathcal N}_i(t)$, and $w_{ij}(t)=0$ otherwise, $\forall i,j=1,\dots,6$. The initial parameter estimates are chosen as $\hat \vartheta_i(0)=[0,0,0]^T$, $i=1,\dots,6$. The joint positions of the robots are shown in Fig. 3 and Fig. 4. The control torques of the robots are shown in Fig. 5 and Fig. 6. We may note that the control torques exhibit switching phenomenon and this is mainly due to the switching of the interaction graph among the robots.

 In the second simulation, we consider the case that there exist communication delays among the robots in addition to the switching topology. The communication delays, for simplicity, are set as $T_{ij}=1 \text{ s}$, $j\in{\mathcal N}_i(t),i=1,\dots,6$. The controller parameters are chosen to be the same as the first simulation. The joint positions of the robots are shown in Fig. 7 and Fig. 8.

\begin{figure}
\centering
\begin{minipage}[t]{1.0\linewidth}
\centering
\includegraphics[width=2.5in]{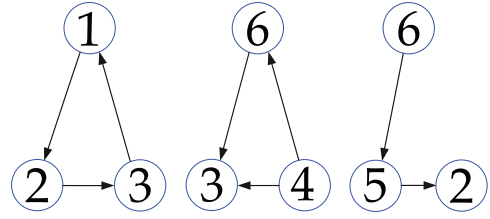}
\caption{Possible interaction graphs among the six robots.}
\end{minipage}%
\end{figure}

\begin{figure}
\centering
\begin{minipage}[t]{1.0\linewidth}
\centering
\includegraphics[width=2.5in]{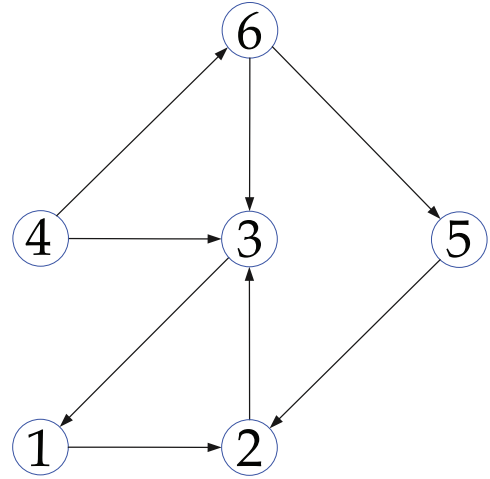}
\caption{Union of the interaction graphs.}
\end{minipage}%
\end{figure}

\begin{figure}
\centering
\begin{minipage}[t]{1.0\linewidth}
\centering
\includegraphics[width=3.0in]{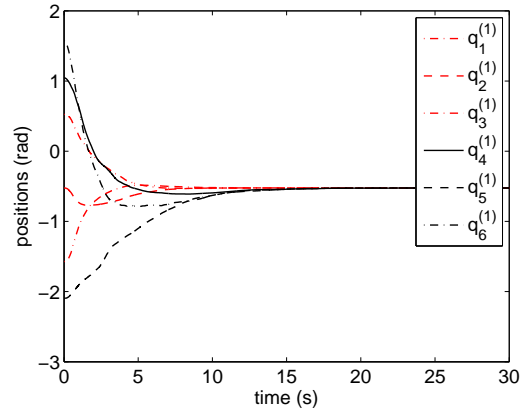}
\caption{Positions of the robots (first coordinate).}
\end{minipage}%
\end{figure}

\begin{figure}
\centering
\begin{minipage}[t]{1.0\linewidth}
\centering
\includegraphics[width=3.0in]{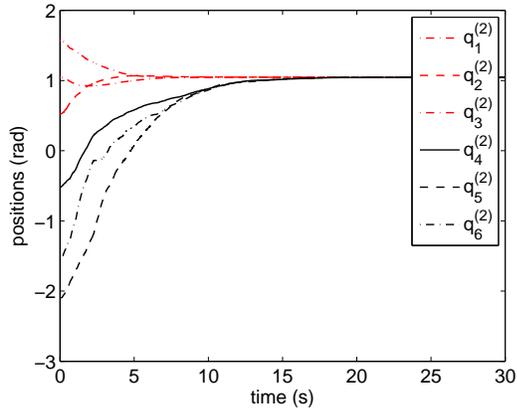}
\caption{Positions of the robots (second coordinate).}
\end{minipage}%
\end{figure}

\begin{figure}
\centering
\begin{minipage}[t]{1.0\linewidth}
\centering
\includegraphics[width=3.0in]{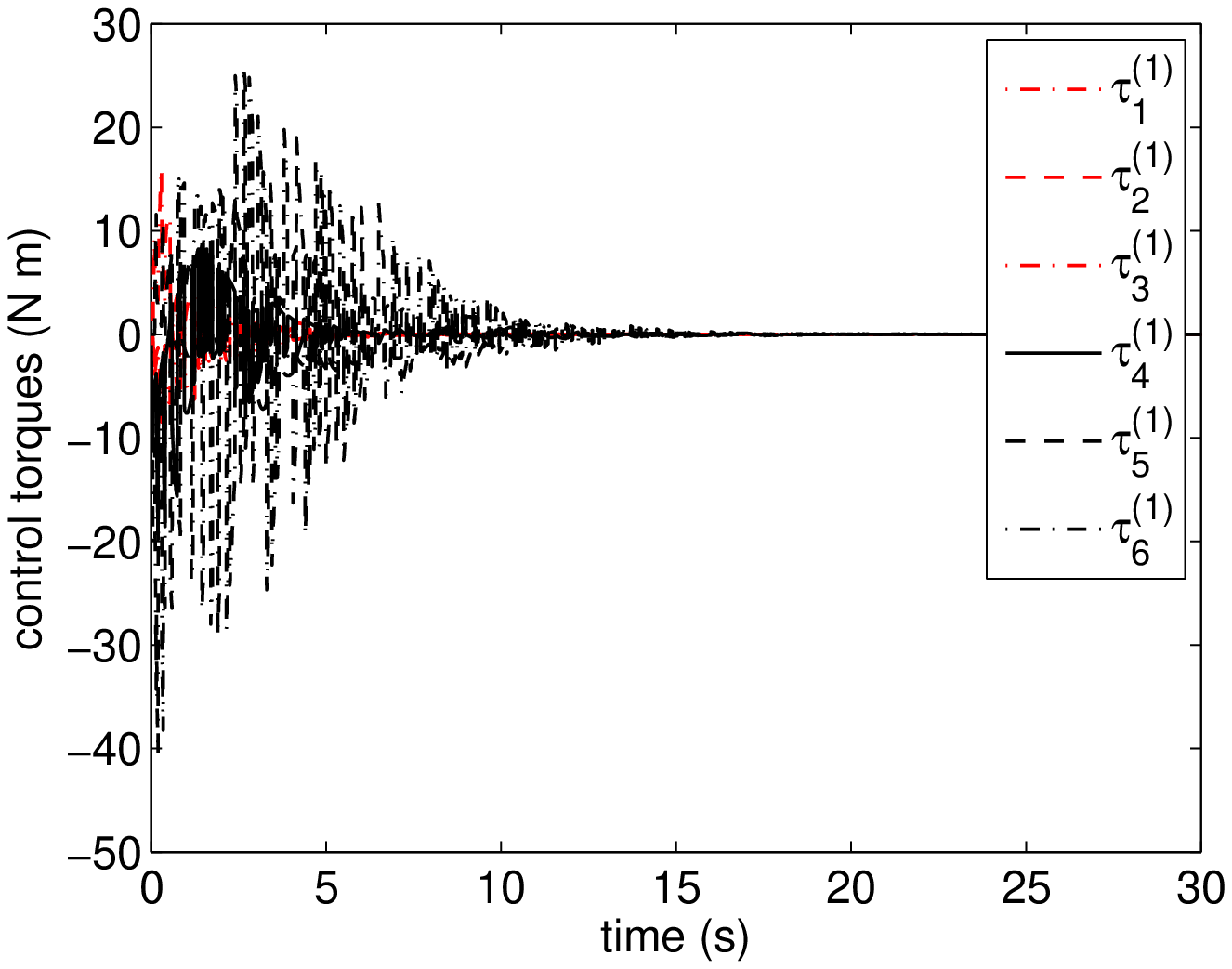}
\caption{Control torques.}
\end{minipage}%
\end{figure}

\begin{figure}
\centering
\begin{minipage}[t]{1.0\linewidth}
\centering
\includegraphics[width=3.0in]{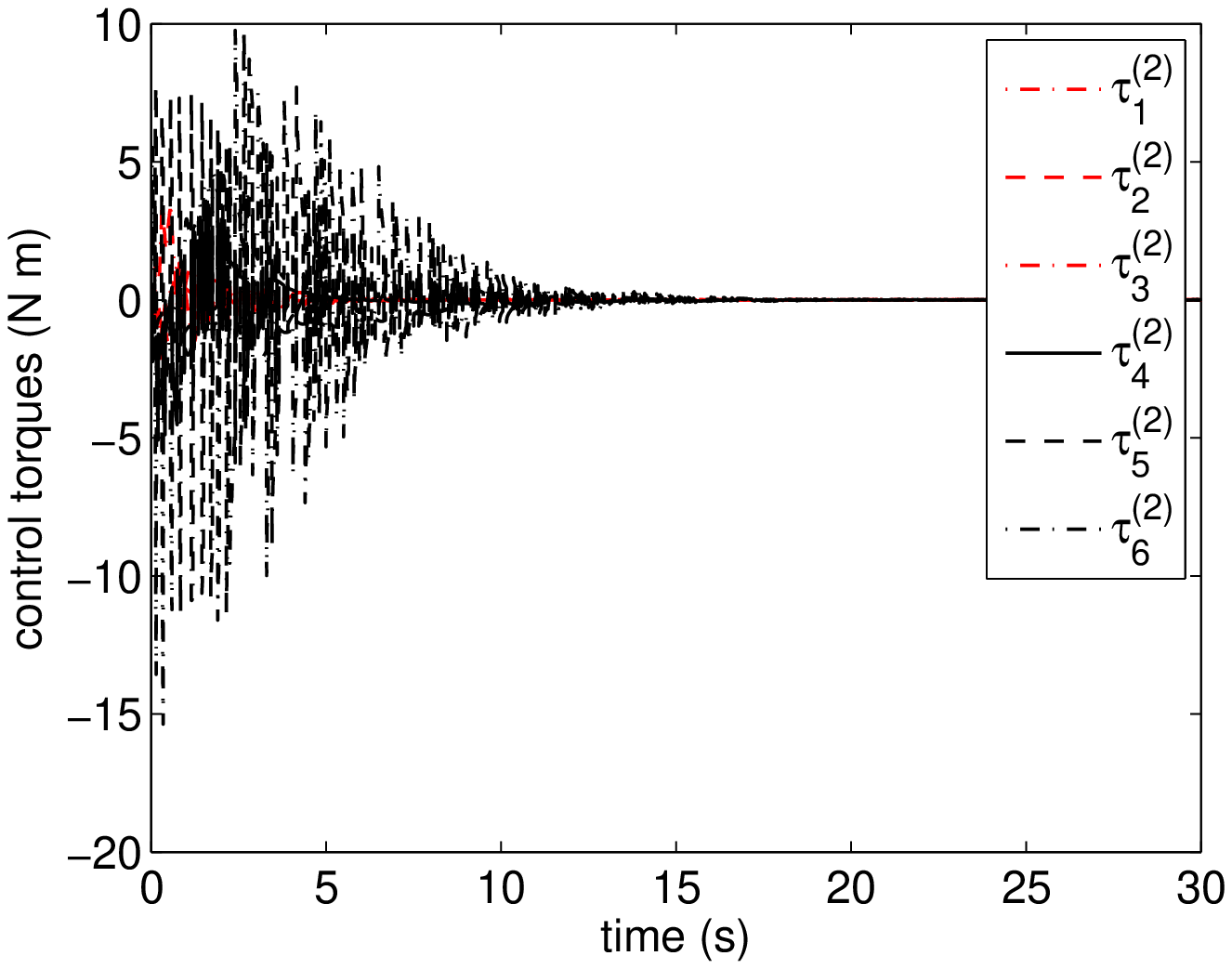}
\caption{Control torques.}
\end{minipage}%
\end{figure}

\begin{figure}
\centering
\begin{minipage}[t]{1.0\linewidth}
\centering
\includegraphics[width=3.0in]{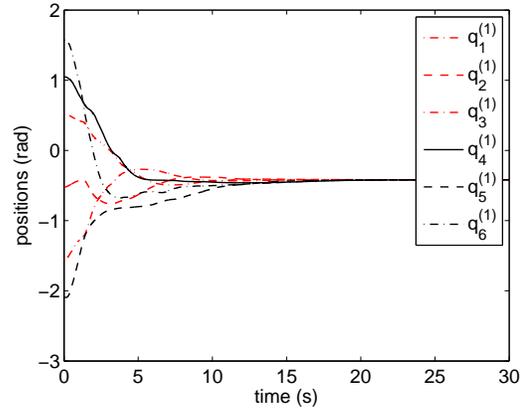}
\caption{Positions of the robots with switching topology and communication delay(first coordinate).}
\end{minipage}%
\end{figure}

\begin{figure}
\centering
\begin{minipage}[t]{1.0\linewidth}
\centering
\includegraphics[width=3.0in]{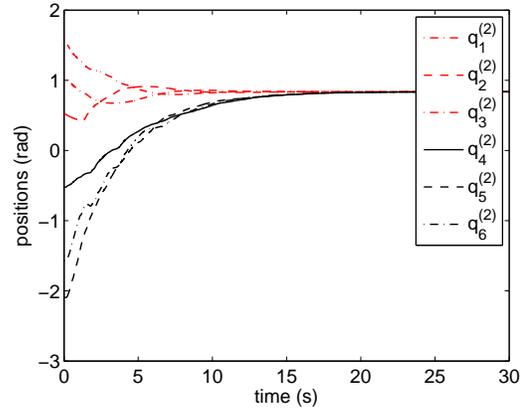}
\caption{Positions of the robots with switching topology and communication delay (second coordinate).}
\end{minipage}%
\end{figure}

%

\section{Conclusion}

In this paper, we have investigated the consensus problem for networked Lagrangian systems. For addressing the discontinuity resulted from the switching topology, a new adaptive controller is developed by employing dynamic feedback and a new analysis tool referred to as unform integral-$\mathcal L_p$ stability is introduced for analyzing the stability and convergence of the networked systems. It is shown that the proposed adaptive controller can ensure that all systems' positions converge to the same value provided that the union of the interaction graphs contains a directed spanning tree, with or without communication delays. Numerical simulations are provided to show the performance of the proposed adaptive controllers.


%








\bibliographystyle{IEEEtran}
\bibliography{..//Reference_list_Wang}

%
%
%

%








\end{document}